\documentclass[twocolumn,aps,prl,superscriptaddress,amssymb,showpacs]{revtex4}
\usepackage[dvips,final]{graphicx}
\begin{document}
\title{Intrinsic Superconductivity at 25~K in Highly Oriented Pyrolytic Graphite}
\author{P. Esquinazi}\email{esquin@physik.uni-leipzig.de}
\affiliation{Abteilung f\"ur Supraleitung und Magnetismus,
Institut f\"ur Experimentelle Physik II, Universit\"{a}t Leipzig,
Linn\'{e}stra{\ss}e 5, D-04103 Leipzig, Germany}
\affiliation{CMAM, Universidad Autonoma de Madrid, Cantoblanco,
E-28049 Madrid, Spain} \affiliation{Laboratorio de F\'isica de
Sistemas Peque\~nos y Nanotecnolog\'ia,
 Consejo Superior de Investigaciones Cient\'ificas, E-28006 Madrid, Spain}
\author{N. Garc\'ia}\email{nicolas.garcia@fsp.csic.es}
\affiliation{Laboratorio de F\'isica de Sistemas Peque\~nos y
Nanotecnolog\'ia,
 Consejo Superior de Investigaciones Cient\'ificas, E-28006 Madrid, Spain}
 \affiliation{Abteilung f\"ur Supraleitung und Magnetismus, Institut
f\"ur Experimentelle Physik II, Universit\"{a}t Leipzig,
Linn\'{e}stra{\ss}e 5, D-04103 Leipzig, Germany}
\author{J. Barzola-Quiquia}
\affiliation{Abteilung f\"ur Supraleitung und Magnetismus,
Institut f\"ur Experimentelle Physik II, Universit\"{a}t Leipzig,
Linn\'{e}stra{\ss}e 5, D-04103 Leipzig, Germany}
\author{J. C. Gonz\'alez}
\author{M. Mu\~noz}
\affiliation{Laboratorio de F\'isica de Sistemas Peque\~nos y
Nanotecnolog\'ia,
 Consejo Superior de Investigaciones Cient\'ificas, E-28006 Madrid, Spain}
\author{P. R\"odiger}
\author{K. Schindler}
\author{J.-L. Yao}
\author{M. Ziese}
\affiliation{Abteilung f\"ur Supraleitung und Magnetismus,
Institut f\"ur Experimentelle Physik II, Universit\"{a}t Leipzig,
Linn\'{e}stra{\ss}e 5, D-04103 Leipzig, Germany}

\begin{abstract}
High resolution magnetoresistance data in highly oriented
pyrolytic graphite thin samples manifest non-homogenous
superconductivity with critical temperature $T_c \sim 25~$K. These
data exhibit: i) hysteretic loops of resistance versus magnetic
field similar to Josephson-coupled grains, ii) quantum  Andreev's
resonances and iii) absence of the Schubnikov-de Haas
oscillations. The results indicate that graphite is a system with
non-percolative superconducting domains  immersed in a
semiconducting-like matrix. As possible origin of the
superconductivity in graphite we discuss interior-gap
superconductivity when two very different electronic masses are
present.
\end{abstract}
\pacs{74.10.+v,74.45.+c,74.78.Na} \maketitle

The standard way to ascribe superconductivity to materials is by
observing the screening of an external applied magnetic field, the
Meissner effect, below a critical field $B_{c1}$ and, although
less important from the physical point of view, by measuring the
drop of resistance to practically zero below a critical
temperature $T_c$. These phenomena are observed for percolative or
homogenous superconductors where a macroscopic wave function of
the Cooper pairs exists \cite{thinkam}. It is well known that in
inhomogeneous superconducting samples, as for example the
well-known ceramic high $T_c$ oxides, sometimes superconductivity
does not percolate, then the resistance does not drop to zero and
the Meissner effect is small. In this case the criteria to assign
non-percolative inhomogeneous superconductivity to a material is
much less obvious. In addition, we would like to discuss here a
superconducting high-$T_c$ material with a very low density of
free electrons or quasiparticles  $n \lesssim 10^{18}~$cm$^{-3}$,
with very different effective masses $m^\star$.  We think that
this is the case of highly oriented pyrolytic graphite (HOPG), the
material studied in this work.

Untreated HOPG samples manifest large electronic mean free path
and Fermi wavelength of order of microns \cite{gon07}. On the
other hand the same samples reveal that the surface is not an
equipotential with metallic and insulating regions that can move
\cite{lu06,gom07}. It seems clear that the view of graphite as a
more or less ordered, homogeneous system  and with a homogeneous
density of carriers cannot be hold and it does not represent the
interesting piece of the physics of HOPG. Although resistance
$R(T)$ data can be fitted, in some cases, with an homogeneous two
band model (TBM) using two mobilities and two carrier
concentrations  (all temperature dependent parameters)
\cite{kelly}, there are other observations as a function of the
applied magnetic field reported here that cannot be explained
within this model. In this work we treat HOPG as a non-uniform
electronic system and as such it will be discussed.

To aboard this hard problem we have obtained over $10^6$ high
resolution magnetoresistance (MR) data points in a range of
temperatures. These data exhibit: (i) irreversible hysteretic
loops of resistance versus magnetic field similar to those
observed in granular superconductors with Josephson-coupled grains
\cite{ji93,kope01} that can be assigned to superconducting
fluxons, (ii) quantum Andreev's resonances  in the MR \cite{gar07}
and (iii) absence of Schubnikov-de Haas (SdH) oscillations. The
experimental data indicate the existence of energy gaps at the
Fermi level and that HOPG is a non-percolative superconductor with
``granular" domains immersed in a semiconductor-like matrix. The
origin of the superconductivity in graphite may be assigned to
interior-gap superconductivity that predicts a gapless stability
when two different masses are present, a problem that has been
discussed by Liu and Wilczeck \cite{liu03}.

The high-resolution, low-noise four-wires MR measurements have
been performed by AC technique (Linear Research LR-700 Bridge with
8 channels LR-720 multiplexer)  with ppm resolution and in some
cases also with a DC technique (Keithley 2182 with 2001
Nanovoltmeter and Keithley 6221 current source). The temperature
stability achieved was $\sim 0.1~$mK and the magnetic field,
always applied normal to the graphene planes, was measured by a
Hall sensor just before and after measuring the resistance, and
located at the same sample holder inside a superconducting-coil
magnetocryostat. We used currents between $1 \ldots 100~\mu$A.

\begin{figure}[]
\begin{center}
\includegraphics[width=85mm]{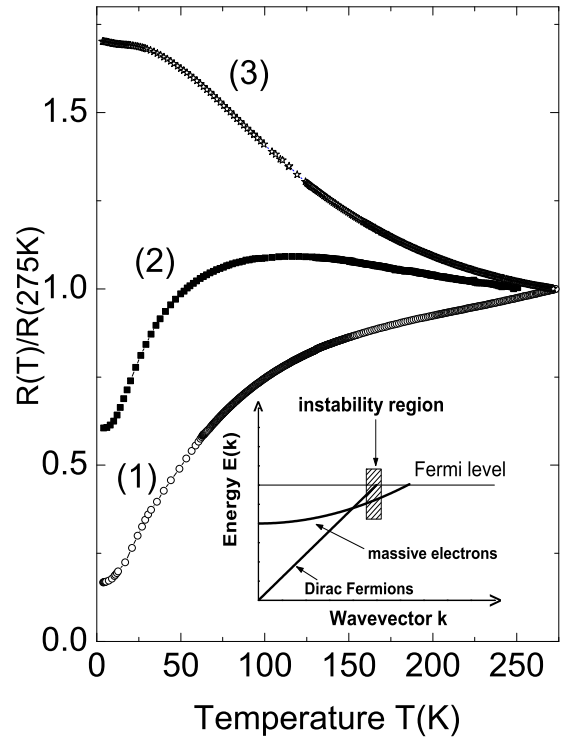}
\caption[]{Normalized resistance as a function of temperature at
zero applied field for three HOPG samples. Samples (1) and (3)
were obtained from the same HOPG grade ZYA ($0.4^\circ$ rocking
curve width) bulk sample from Advanced Ceramics. Particle induced
x-ray emission indicates impurity levels of metallic elements
below $5~\mu$g/g with exception of V ($16~\mu$g/g). Sample (1)
$(R(275~$K) = 72~m$\Omega$) was $\sim 10~\mu$m thick and $4.4$~mm
length. Sample (3) $(R(275~$K) = 16~$\Omega$) was 12~nm thick and
$\sim 30~\mu$m length and width, with $\sim 9~\mu$m distance
between nearby electrodes, see inset in Fig.~\protect\ref{osc}(a).
The Pd-electrodes (to avoid Schottky barriers) were prepared using
conventional electron lithography. The sample (2) $(R(275~$K) =
6.5~m$\Omega$) was obtained from HOPG grade ZYC bulk sample
($3.5^\circ$). The inset shows schematically the energy dispersion
relations for two carriers in graphite, massive and massless
(Dirac fermions). Following Ref.~\protect\onlinecite{liu03}, the
instability region lies around the Fermi wavevector of the light
particles.} \label{rt}
\end{center}
\end{figure}

To start with our strategy we have prepared different samples of
HPOG that just differ in its ordering and size and they exhibit
apparently different behaviors with T. Figure \ref{rt} shows
$R(T)$ for the samples indicated in the figure caption. Usually
one tends to fit these curves with the TBM. In particular the
$R(T)$ of sample (3)  can be fitted approximately. However,
carriers in HOPG have two different masses and one of them is
practically zero corresponding to Dirac electrons \cite{luky04}.
Furthermore, there are other important aspects described below
that undoubtedly cannot be put into accord with the TBM. We
concentrate in the very thin and micrometer small sample because
it should have less number of fluctuating domains and this should
provide more clear superconducting-related effects. Note that this
sample shows a semiconducting like behavior that levels off at $T
\simeq 25~$K; its in-plane resistivity $\rho_{ab} (10~$K$) \simeq
(50 \pm 10)~\mu\Omega$cm is similar to the one of sample (1) from
which it has been obtained by careful exfoliation.

\begin{figure}[]
\begin{center}
\includegraphics[width=85mm]{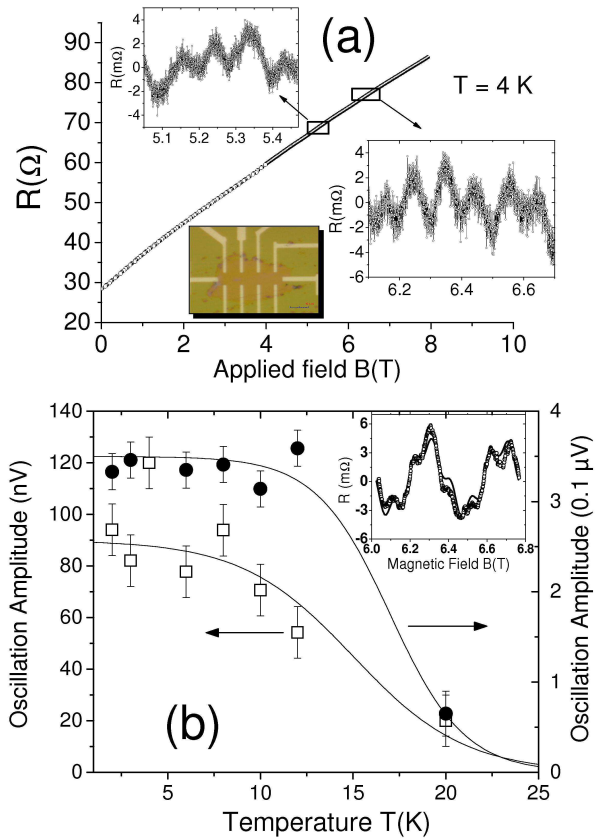}
\caption[]{(a) Resistance of sample (3) between two adjacent
voltage electrodes as a function of magnetic field. A close
inspection of the MR of this sample at fields above 0.5~T reveals
an anomalous behavior, namely the MR oscillates. The oscillations
shown in the insets were obtained after subtracting a quadratic
field dependence around 5.3~T and 6.4~T. These small-field-period
oscillations in the resistance are superposed to oscillations of
larger amplitude and field period, see inset in (b). Further
measurements indicate that the overall shape, field positions and
period are independent of the field sweeping rate, field step and
field sweep direction. The oscillations are observed at low as
well as high fields, as expected because the slope of $R$ vs. $B$
does not depend appreciable with field \protect\cite{gar07}.
Different periods as well as oscillation amplitudes are observed
for other samples. The inset shows an optical microscope picture
of the sample with the Pd-electrodes. Counting clockwise from
input current electrode 1 at the right, the data shown were taken
between electrodes 3 and 4 (Ch.2). (b) Temperature dependence of
the voltage amplitude of the two oscillations taken from the
Fourier fit, see inset. The continuous lines are a guide. The
inset shows the data at 2~K after subtraction of a linear field
background and the continuous line is the Fourier fit with periods
0.1~T and 0.387~T. These periods are independent of temperature
within experimental resolution.} \label{osc}
\end{center}
\end{figure}

Figure~\ref{osc}(a) shows the MR of sample (3) at 4~K in detail
and in the region 4~T to 8~T with larger resolution using a
magnetic field step of $\simeq 1$~Oe. The first surprise is that
the MR is very small compared with the MR of larger samples of
HOPG. In these samples the ordinary MR of HOPG between 0~T and 8~T
is $\sim 10000\%$ while in the small sample measured here is only
$\lesssim 300\%$. This difference is discussed in
Ref.~\onlinecite{gon07}. In addition, SdH oscillations are absent
in sample~(3) (in other samples of similar size we measured they
appear very weak). This might imply that the Fermi level lies in a
gap. Notice that we decided to perform experiments with very small
field increment. This was not done accidentally. The reason is
that we expected to have weak quantum oscillation resonances --
compared with the classical SdH oscillations -- due to the small
number of potential fluctuations (note that the sample is small,
of the order or smaller than the mean free path and Fermi wave
length) and these fluctuations will induce an oscillating
transmissivity through the potential wells. These quantum
oscillations were proposed theoretically to interpret observed
structures that were over seen or consider noise in graphene
samples \cite{gar07}. And of course the sample of Fig.~\ref{osc}
shows the expected quantum oscillations. These quantum
oscillations have a two period spectrum indicating that in the
sample one has at least two characteristic potential wells.
Figure~\ref{osc}(b) shows the oscillation amplitude of the two
harmonics  (see also the inset) as a function of $T$, which remain
constant below 10~K and vanish at a critical temperature $T_c
\simeq 25~$K.

We claim that these oscillations, given their small amplitude of
$\sim 100~$nV$ \ldots 400~$nV (much smaller than the corresponding
values in temperature for the used range $T \eqslantgtr 2~$K) are
due to the interference of wave functions that suffer Andreev's
reflections at the potential walls matching low-gap semiconducting
with superconducting regions.  From the period of the oscillations
in field we can estimate that there are superconducting
``granular" domains of size around $1~\mu$m separated by small-gap
semiconducting matrix of similar size, which  couples the
superconducting grains. If this picture is realized one expects to
see pinning and dissipation effects due to fluxons, as discussed
by Ji et al. in Ref.~\onlinecite{ji93}, with circumvent
superconducting currents between the superconducting grains
through the semiconducting regions. One may argue against the
physical ground of the model we are proposing: how is it possible
that superconducting pairs can be kept in a micron-size
semiconducting-like regions connecting the superconducting ones?
This should not be a problem. By using nano-fabricated
constrictions and measuring the transition from ohmic to ballistic
transport we have observed that the mean free path of the carriers
in HOPG at 10~K is $\gtrsim 10~\mu$m. Therefore, it should be
perfectly possible that the pairs travel $\sim 1~\mu$m distance
without breaking out. In other words the proximity effect in
graphite may extend to microns.

\begin{figure}[]
\begin{center}
\includegraphics[width=85mm]{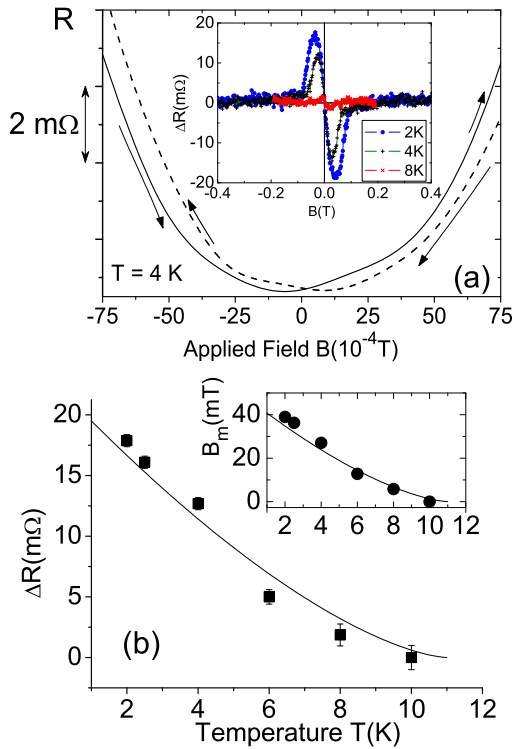}
\caption[]{(a) Strongly enhanced MR curve near zero field. A weak
hysteresis appears similar to butterfly MR loops for
superconductors with Josephson-coupled grains \protect\cite{ji93}.
For a clear observation of the hysteresis we present in the inset
the difference of the resistance curves (see text).  (b) The
height of the irreversibility maximum $\Delta R$ as well as their
field positions $B_m$ (see inset) vs. temperature. The continuous
lines follow the function $\propto (1 - (T/11))^{1.5}$.}
\label{irr}
\end{center}
\end{figure}

If there are fluxons then one should have irreversible hysteretic
loops of the kind observed in granular superconductors
\cite{ji93,kope01}. Figure~\ref{irr}(a) shows this irreversibility
that cannot be explained by ferromagnetism, ferroelectricity due
to motion of charges or by usual Abrikosov vortices, since no sign
of irreversibility has been seen within experimental error for
 magnetic fields applied parallel to the planes.
We have a huge anisotropy in an otherwise a small spin-orbit
coupling material. Note that the two minima in $R$ are observed at
the positive and negative fields coming from high fields from the
same direction. Only by fluxons running between the
superconducting and the semiconducting-like regions these
hysteresis loops can be explained. For a better appreciation of
the hysteresis the inset in Fig.~\ref{irr}(a) shows the difference
between the two curves, i.e.~the resistance curve obtained by
starting at a negative field and sweeping to positive fields is
subtracted from the resistance curve measured when starting at a
positive field and sweeping to negative fields. The height of the
extreme as well as their fields $B_m(T)$ depend on $T$. The
$T$-dependence of this irreversibility $\Delta R$ as well as
$B_m(T)$ vanish at $T_{i} \sim 11~$K. The reason why the
irreversible behaviour shown in Fig.~\ref{irr} vanishes at $\sim
11$~K in contrast to the $\sim 25$~K observed from the oscillatory
behavior of Fig.~\ref{osc}, can be easily related to the pinning
of the fluxons inside the grains. The temperature dependence of
the irreversibility in field, continuous lines in
Fig.~\ref{irr}(b), follows $(1-(T/T_i))^{1.5}$ a similar
dependence as for the irreversibility line of vortices observed in
high-temperature superconductors.

Because it is just graphite, the superconducting regions have a
very small number of free electrons,  say $\lesssim 10^{-4}$
electrons per carbon atom \cite{kelly}. A simple estimate shows
that the London penetration length is larger than microns and
therefore the Meissner effect should be unnoticeable. Also the
resistance does not drop to zero because the superconducting
regions do not percolate, in additions to the resistance due to
the motion of fluxons.   The observed hysteresis is a very strong
fingerprint of superconducting fluxons, difficult to rule out.

The density of carriers in HOPG samples is very probably highly
inhomogeneous, and upon region in the sample it may be much
smaller than $10^{-4}$/C-atom.  What might be the physical origin
of this superconductivity? Graphite contains two carrier families
with very different $m^{\star}$, one with a negligible mass called
Dirac fermions. Therefore, the ratio between masses may be very
large, 100 or larger. These different masses establish large
instabilities if the number of the different carriers is not that
different. Then we have a large extended band with a large Fermi
energy corresponding to the light carriers and a lower Fermi
energy for the heavy carriers, see inset in Fig.~\ref{rt}. A large
density of the heavy carriers pinned at the Fermi energy of the
light particles has strong electron interaction and creates
instabilities that will be discussed in other work. In particular
for this situation Liu and Wilczek \cite{liu03} have predicted a
condensed superfluid state called interior-gap superconductivity
or breached superconductivity. Graphite might be a good candidate
where some concepts of this theory could be useful. In fact the
picture they describe for their theory \cite{liu03} is similar to
that of the inset in Fig.~\ref{rt}. In this theory no gap exists
and the material may exhibit $p$-type superconductivity, which has
been also discussed for graphite \cite{gon01} as a more robust
state in a non-homogeneous system.

The results of Figs.~\ref{osc} and \ref{irr} belong to a
micrometer size sample (parallel to the planes) and 12~nm
thickness in order to have few potential fluctuations.
Measurements in two other samples of similar size show similar
behavior but slightly different $T_c$'s. In larger samples, as for
example the other two reported in Fig.~1, the same type of effects
should be seen but more in terms of universal conductance
fluctuations. In fact we have observed in these and other larger
samples fluctuations in the resistance up to room temperature,
however they are difficult to tackle down and their amplitudes
change with time, an effect that is probably related to the motion
of charges with current and applied magnetic field.
Superconductivity in graphite should by no means limited to the
25~K here obtained for the small sample, but depends on the charge
density, defect density and the related instabilities at Fermi
level.

We note that hints for superconductivity in HOPG samples
 from SQUID measurements have been invoked in the past
\cite{kopejltp07}. However, resolution limits of the magnetometer
and the partial admixture of ferromagnetic-like signals casted
doubts on the origin of those signals. Other studies
\cite{yakovadv03} claimed
 superconductivity in graphite based on the metal-insulator transition
  observed under a magnetic
 field, although superconductivity does not necessarily need to
 be invoked to understand this transition. There is also a theoretical work
  that claims high-$T_c$ $d$-wave superconductivity
 in graphite based on resonating valence bonds \cite{doni07}.

Concluding, in this work we have obtained evidence that supports
the existence of intrinsic superconductivity in HOPG based on the
irreversibility of the MR and on the quantum oscillations. We
think that interior-gap -- breached superconductivity \cite{liu03}
is an interesting starting concept to understand the observed as
well as other phenomena in the transport properties of graphite.

We gratefully thank Y.Kopelevich  for fruitful discussions on the
superconductivity of graphite. This work was done with the support
of the DFG under ES 86/11, the Spanish CACyT and Ministerio de
Educaci\'on y Ciencia. J.-L. Yao acknowledges the support from the
A. von Humboldt foundation.


\end{document}